\documentclass[conference,a4paper]{IEEEtran}
\usepackage[utf8]{inputenc}

\usepackage{microtype}
\usepackage[font=small,labelfont=bf]{caption}
\usepackage{siunitx,textcomp}

\makeatletter

\usepackage[backend=biber,style=ieee,sorting=none,giveninits=true]{biblatex}
\usepackage[usenames,dvipsnames]{xcolor}
\usepackage[textsize=small]{todonotes}
\makeatletter
\g@addto@macro{\UrlBreaks}{\UrlOrds}
\makeatother
\newcommand{\note}[1]{
}

\author{\IEEEauthorblockN{
Jonathan Vestin\IEEEauthorrefmark{1}, Andreas Kassler\IEEEauthorrefmark{1}, Deval Bhamare\IEEEauthorrefmark{1}, \\
Karl-Johan Grinnemo\IEEEauthorrefmark{1}, Jan-Olof Andersson\IEEEauthorrefmark{1}, Gergely Pongracz\IEEEauthorrefmark{2}}
\vspace{-2ex}
\\
\IEEEauthorblockA{
\IEEEauthorrefmark{1}Karlstad University, Sweden, 
\IEEEauthorrefmark{2}Ericsson AB, Hungary 
\vspace{-2ex}
\\\\
\{jonathan.vestin, andreas.kassler, deval.bhamare, karl-johan.grinnemo\}@kau.se\\janoande02@student.kau.se \\
\vspace{-2ex}
gergely.pongracz@ericsson.com}}
\title{Programmable Event Detection for In-Band Network Telemetry}
\date{June 2019}

\global\csname @topnum\endcsname 0
\global\csname @botnum\endcsname 0

\begin{document}
%\IEEEoverridecommandlockouts
%\IEEEpubid{\makebox[\columnwidth]{978-1-7281-4832-8/19/\$31.00~\copyright2019 IEEE \hfill} \hspace{\columnsep}\makebox[\columnwidth]{ }}

\maketitle

% \IEEEpubidadjcol

\begin{abstract}
    In-Band Network Telemetry (INT) is a novel framework for collecting telemetry items and switch internal state information from the data plane at line rate. With the support of programmable data planes and programming language P4, switches parse telemetry instruction headers and determine which telemetry items to attach using custom metadata. At the network edge, telemetry information is removed and the original packets are forwarded while telemetry reports are sent to a distributed stream processor for further processing by a network monitoring platform. In order to avoid excessive load on the stream processor, telemetry items should not be sent for each individual packet but rather when certain events are triggered. In this paper, we develop a programmable INT event detection mechanism in P4 that allows customization of which events to report to the monitoring system, on a per-flow basis, from the control plane. At the stream processor, we implement a fast INT report collector using the kernel bypass technique AF\_XDP, which parses telemetry reports and streams them to a distributed Kafka cluster, which can apply machine learning, visualization and further monitoring tasks. In our evaluation, we use real-world traces from different data center workloads and show that our approach is highly scalable and significantly reduces the network overhead and stream processor load due to effective event pre-filtering inside the switch data plane. While the INT report collector can process around 3~Mpps telemetry reports per core, using event pre-filtering increases the capacity by 10-15x. 
    
\end{abstract}

\section{Introduction}
Operations, Administration, and Management~(OAM) refers to protocols, tools and mechanisms that help network operators in fault indication, performance monitoring, security management, diagnostic functions, accounting, performance monitoring, configuration and service provisioning. In traditional carrier networks, OAM tools such as SNMP and OWAMP-Test are used, however, these tools have been proven inadequate for SDN-NFV data centers. They are not  scalable and cannot provide  fine-grained, real-time information about the overall performance of the data center infrastructure~\cite{8406169}. 

In-band Network Telemetry~(INT) has gained a lot of momentum over the last few years~\cite{8406169,burstradar,kim2015band,sINT, INT-OPT}. The idea behind the INT framework is that each node along a network path adds telemetry items and network state to in-band, data plane traffic. Telemetry items may include switch ID, ingress timestamps, queue occupancy information, and various other performance-related metadata, which are added at line rate as customized headers to in-band, data plane packets.  The telemetry items are forwarded to a distributed network monitoring platform, which uses stream processors to consume the telemetry metadata, compile traffic statistics, and apply machine learning to derive actions and/or recommendations for the operation and management of the data center. Eventually, the original data plane packets are recovered at the network edge and forwarded to the end-user.

The advent of programmable data planes and high-level, platform-independent programming languages such as P4~\cite{bosshart14}, have enabled fine-granular monitoring of data plane traffic. \cite{cugini2019p4} uses P4-based INT to collect per-packet queue statistics in a metro network, and use that information to enforce QoS; \cite{niu2019visualize} proposes such a telemetry solution for IP-over-optical networks; and Barefoot Networks includes their INT technology in the Smart Programmable Real-time In-band Network Telemetry (SPRINT). A common denominator to these and other previous work is that it lack highly configurable per-flow event detection in the data plane. This results in high load on INT stream processors and limits the scalability or severely limits the event collection possibilities.

In this paper, we make the following contributions. First, we design a programmable event detection framework for INT data. The control plane is responsible to specify high-level event algorithms along with threshold values that trigger the generation of telemetry reports from the programmable data plane. Second, we implement several event detection and filtering algorithms in P4 for INT metadata. Third,  we develop an INT monitor using kernel bypass AF\_XDP~\cite{toke18}, which allows to parse large amounts of INT reports per second on a single core and forward them to a Kafka~\cite{kafka} cluster for further processing. The SDN controller can use the telemetry reports and fine-tune the event detection mechanisms on a per-flow basis according to, e.g., service level agreements and detected events.
Our evaluation shows that a single core can process \SI{3.593}{Mpps} telemetry reports without P4 event detection. When using  event detection, using a filter threshold of 100~$\mu$s queue buildup for \emph{web} traffic increases that to \SI{35.22}{Mpps}, while for \emph{database} traffic it increases to \SI{12.21}{Mbps}. 
 %To this end, our solution features per-flow, threshold filtering, avoids excessive load on the INT Collectors, and thus becomes more versatile and scalable than most existing INT solutions.
 
The rest of the paper is organized as follows. Section~II discusses the INT framework and related work. Section~III describes the design and implementation of our solution. Section~IV evaluates our P4 design in a testbed and Section~V concludes the paper.

\section{Background and Related Work}
%\todo[inline]{Commented subsection A%If we have to split this section, I suggest to move the text of sub-section A in the 2nd paragraph of Introduction, as there is some repetition. It will strengthen the introduction. Also it will save some space.
%}

%%In-Band Network Telemetry (INT)~\cite{kim2015band, INT-OPT} is  a new network monitoring framework which enables  data plane devices to collect telemetry information at line rate. With the emergence of programmable data plane elements and programming language support such as P4, telemetry instructions can be attached to individual packets, which the P4 programmable switch would parse and understand, in order to collect desired telemetry items. Such monitoring information may include which path the packet took, queue occupancy, packet latency experienced at switch, packet time-stamp per switch, etc. and is added as custom headers for each packet. Using the INT framework, user data packets thus collect observed network state from INT enabled devices while traversing the network and aggregate such state in flexible headers that can sent as telemetry reports to monitoring devices in order to infer hop-by-hop or end-to-end latency~\cite{kim2015band},~\cite{lapukhov-dataplane-probe-01}.  

%%\subsection{Related Work}
Both industry and academia have explored the area of network monitoring, its overhead on network infrastructure, and how to minimize it \cite{bhamare2015models}  However,  existing  solutions  mainly  focus  on  the trade-off  between  expressiveness,  accuracy and speed, and often stress the monitoring platform. For example, systems such as NetQRE~\cite{yuan2017quantitative} can  support  a  wide  range  of  queries  using  stream processors  running  on general-purpose  CPUs,  but  they  incur  substantial  bandwidth  and  processing  costs  to do  so. Telemetry systems such as Chimera~\cite{borders2012chimera}
and Gigascope~\cite{gigascope} are expressive  in  nature  by  covering  a wide  range  of  telemetry items,  however,  can  only  support  lower  packet rates.  This is  because  these  systems  process  all  packets  at  the  stream processor which can become a bottleneck.

Several works try to reduce the amount of INT information that INT monitors and stream processors need to process. Clearly, there is a trade-off between the accuracy of the captured network state and the load imposed on the monitoring framework. Sonata~\cite{gupta18} is a framework for performing complex in-network event detection, through an approach which splits event processing between user-space stream processors and programmable switch data planes. Through the use of simple, but powerful well-known operators like \emph{filter}, \emph{map}, \emph{reduce} etc. %Sonata provides a declarative language for specifying telemetry tasks. 
INTCollector~\cite{tu18} implements a distributed telemetry monitoring system by parsing telemetry reports inside the Linux kernel using eBPF and XDP. It implements threshold- and interval-based event detection at the telemetry collector in the fast path and inserts them into a distributed database (e.g., Prometheus\footnote{\url{https://prometheus.io}} or InfluxDB\footnote{\url{https://www.influxdata.com/}}) in the slow path. Because of its single-core implementation it does not scale adequately and its processing rate is limited to 1.2 Mpps for a single core. Because the events are detected only at the telemetry monitoring system, unnecessary telemetry reports are sent from the data plane resulting in a very high load on the INTCollector. 

%%UNCOMMENT IF SPACE This language unifies both stream processor and data plane programming, and compiles into a program that is split between the data plane fast path, and the stream processor slow path. This allows simple, fast operations to be performed in the data plane, while complex operations, or operations not supported by specific data plane target platforms, are offloaded to the stream processor.%% 

%From sonata: 
%These are Datapath variants of Sonata: Marple, OpenSketch, Everflow, Unimon
%Telemetry Papers: Chimera, Gigascope, OpenSOC, NetQRE
%Can we find any other relate
Joshi et al.~\cite{burstradar} present BurstRadar, where the key idea is to first detect a micro burst in the data plane and then capture a snapshot of telemetry information of all the involved packets. This information allows queue composition analysis to identify the culprit flow(s), and burst profiling to know burst characteristics such as duration, queue build-up/drain rates, etc. Authors in~\cite{chen2018catching} also propose an approach called Snappy to to identify the particular flows responsible for a micro burst, and handle them automatically. Snappy maintains multiple snapshots of the occupants of the queue over time, where each snapshot is a compact data structure that makes efficient use of data-plane memory. 

Kim, Suh and Pack~\cite{sINT} propose selective INT (sINT) where the ratio of packets to be monitored can be adjusted depending on the frequency of significant changes in network information. Li et al.~\cite{li2016flowradar} propose FlowRadar, a novel approach to maintain flows and their counters that scale to a large number of flows with small memory and bandwidth overhead. The key idea of FlowRadar is to encode per-flow counters with a small memory and constant insertion time at switches, and then to leverage the computing power at the remote collector to perform network-wide decoding and analysis of the flow counters. N. Tu et al. \cite{8094182} have presented an INT architecture for the UDP and discuss its design and integration with Open Network Operating System (ONOS) controller.
%\todo[inline]{Don't understand "...discuss its design...in Open...} 

In our proposed approach, we have modified the INT Collector to offload the event detection from the stream processor to an in-network P4 application. Not only does this reduce the network overhead, it also reduces the stream processor load, since it enables pre-filtering of telemetry reports inside the data plane. 

\section{Design and Implementation}

\begin{figure}
    \includegraphics[width=\linewidth]{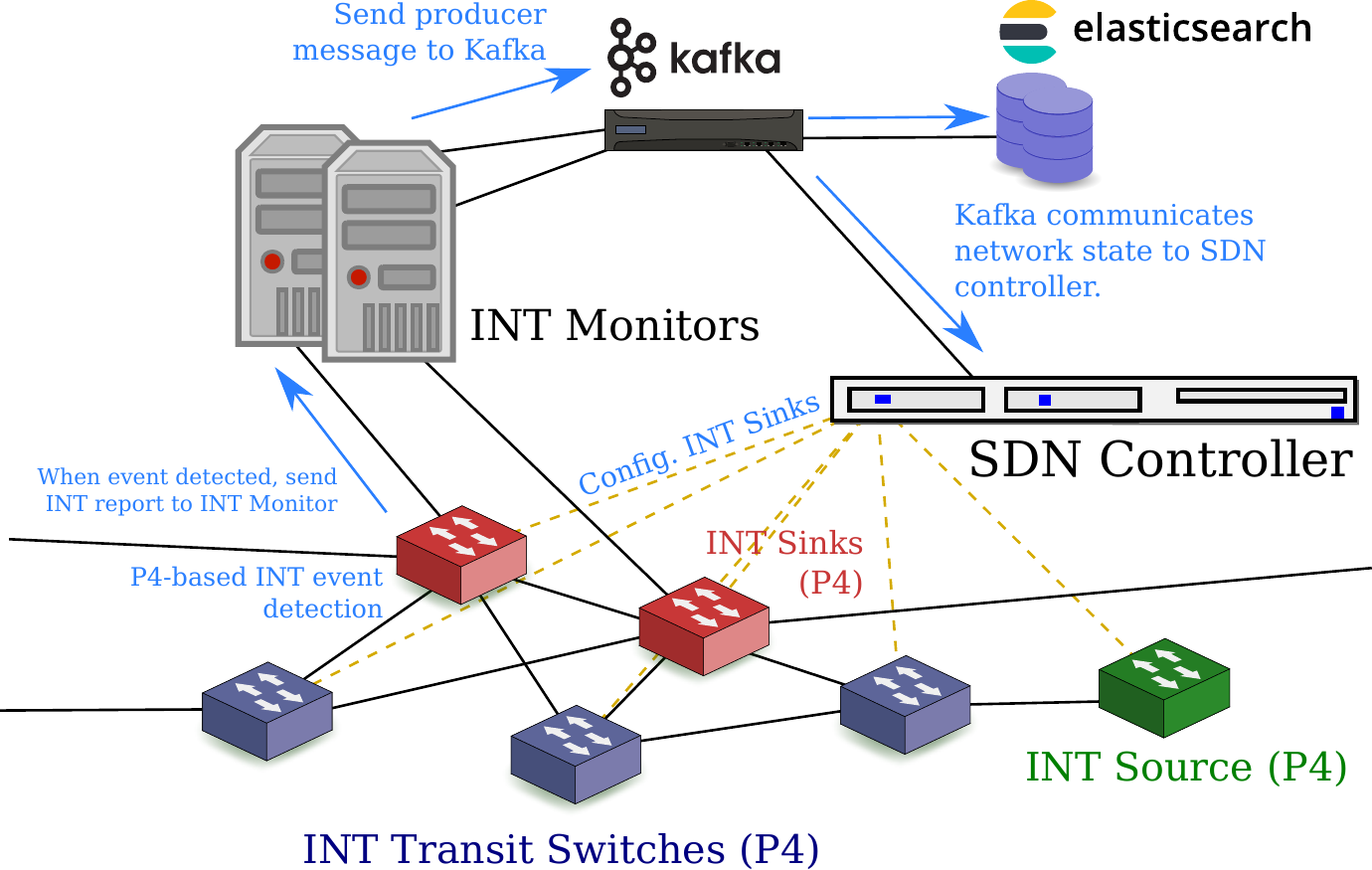}
    \caption{Overview of the INT monitor application}
    \label{fig:int-monitor}
\end{figure}

Network monitoring in data center networks must balance fine-granular monitoring with expressiveness in order to detect events that are important for network operators to act upon swiftly. Especially, high configurability, scalability and line-rate processing are important design goals. The detection of important events occurring in data center network infrastructure, such as micro bursts, may require correlation of multiple stateful variables. These events are of high interest to network operators, as well as customers, as they may negatively impact important service flows. For example, an event may be defined as when the queue occupancy of any switch in a forwarding path is greater than $50\%$ while the end-to-end latency is higher than $5ms$. On the contrary, micro bursts may not be important for a service flow if end-to-end latency is lower than for example $3ms$. Although complex event processing in the data plane on application layer data has been previously explored~\cite{kohler18,vestin18}, this paper applies it to network telemetry data. 

\subsection{Overall Design}
In our design, we allow specifying per flow event thresholds, thus minimizing the load on the stream processors. This is important as in the INT framework, typically telemetry information is added to each user data packet. Sending telemetry reports for each packet to the network monitoring system results in excessive load on the stream processor. Some of that telemetry information can be filtered out easily, because it does not provide more information. For example, when two packets of the same service flow are processed back-to-back in a switch, the queue latency may not have changed significantly and thus the  second packet's telemetry information for queue latency could be discarded.

In our approach, the control plane, through an SDN controller, configures event detection mechanisms, including switch role (e.g. source, sink, transit), detection algorithm and threshold values on a per-flow basis (see Figure~\ref{fig:int-monitor}). This configuration is communicated to the switch through the controller API exposed by P4 (using  table entries and registers). Through this configuration API, the SDN controller specifies the event detection parameters which are executed at the INT sink, allowing for the detection of many different types of events (e.g. queue or latency buildup). Due to the per-flow granularity, this also allows for capturing bi-directional flows.

INT sources insert telemetry headers and an instruction bitmask that determines which telemetry items to collect. When an INT transit switch receives a packet with a telemetry header, it parses the instruction bitmask and pushes INT metadata (e.g., queue occupancy or hop latency) into that header. Finally, when an INT sink receives a packet with a telemetry header, it removes all telemetry headers, and runs the event detection algorithm as configured from the control plane on the appended metadata. Should the criteria for an event be met (e.g., an end-to-end latency spike above the threshold or  queue occupancy over the configured threshold), a telemetry report is sent to the INT monitors (stream processor). In case there is no event detected (e.g. latency below configured threshold), the sink discards the telemetry items thus reducing the load on the stream processor.The INT monitor takes incoming telemetry reports and creates Kafka producer messages, which then can be sent to a Kafka topic. 

The SDN controller can use this data to further reconfigure and fine-tune the threshold and algorithm settings on a per-flow basis; increasing the report resolution for interesting traffic, while reducing it for, e.g., background traffic. As thresholds and telemetry items can be configured per flow basis, our approach also supports customized per slice INT monitoring. Kafka also communicates the INT event reports to an Elastic Search stack for further analytic processing and  visualization. 

\subsection{High Performance INT Collection in AF\_XDP}
For scalable processing of INT telemetry reports, we implemented an INT monitor in~C\footnote{Code available here: \url{https://github.com/jonavest/int-afxdp-kafka}}, using the AF\_XDP~\cite{toke18} socket type for kernel bypass. It also uses the librdkafka\footnote{\url{https://github.com/edenhill/librdkafka}} library for Kafka producer message generation. When launched, it is associated to a specific queue on a network interface, and uses up a single CPU core from the host machine. Multiple instances of the same program can be run simultaneously to enable multi-core processing. 
Our implementation follows the telemetry report format specification given in \cite{intspec}, which includes 8 metadata fields: \emph{switch ID}, \emph{hop latency}, \emph{queue occupancy}, \emph{ingress timestamp}, \emph{egress port id}, \emph{queue congestion status}, \emph{egress port utilization}. While some of the metadata items, like \emph{switch ID}, are uninteresting to run event detection on, we still support parsing them in INT monitor, and we also parse the reserved INT fields which may be used in the future, or by non-standard network extensions. 

\subsection{P4 based Event Pre-filtering using Programmable Data Planes}
Our P4 implementation of the INT framework follows the telemetry report format specification for event detection. To configure the switch, it exposes a series of parameters to the SDN controller. Some of these are configured using the P4 register API, while others are configured using flow tables: 

\begin{LaTeXdescription}
    \item[INT~Mode] The mode which the switch is running for the given flow, which can be either: INT sink, INT transit or INT source. 
    \item[Instruction~mask] When the switch is in the INT source mode, this configures on a per-flow basis which instruction bitmask should be appended to incoming packets. This is configured through a match-action table matching on a flow match key (e.g. mac src/dst, ip src/dst, port number). 
    \item[Algorithm] It is possible to configure the event-detection algorithm and its parameters (such as threshold, metadata type, although algorithms may differ in the number and type of parameters). Also configured through a match-action table with the \emph{fast\_detection} action. 
    \item[Expression] If more complex event detection is required, logic tables installed through P4 registers in conjunctive normal form, adopted from FastReact~\cite{vestin18}, can be used. This uses the \emph{complex\_detection} in the algorithm match-action table. Using complex expression trades performance for allowing more complex expressions. 
\end{LaTeXdescription}

The processing of incoming packets is divided into parsing, ingress processing and egress processing. The parsing consists of a parsing tree, which parses incoming messages looking for an INT header. If an INT header is found, the switch parses any INT metadata available. In the ingress processing, the \emph{ingress timestamp} is recorded into the packet INT metadata, and a regular IPv4 switch forwarding table is applied. In the egress block, we start by fetching the algorithm ID and parameters from the algorithm match-action table. If the flow is configured with \emph{fast\_detection}, the switch runs one out of three algorithms, specified by parameters given to the \emph{fast\_detection} action: \emph{per-hop}, \emph{per-flow} or \emph{moving average}. Also, the INT metadata type (e.g., queue occupancy or hop latency) which the algorithm is run on, is also specified as a parameter. 

The \emph{per-hop} algorithm keeps a record of the previous INT metadata for each \emph{switch~id} in a P4 register table. When a telemetry item is processed by the algorithm, for each \emph{switch~id}, it looks at the difference between the incoming INT metadata and the previous metadata. If the difference exceeds the configured threshold, it stores the new value in the register and marks the packet as an event. This is done through an unrolled loop. The \emph{per-flow} algorithm on the other hand, cares for the sum of the INT metadata for all hops in a specific flow. It stores a table, for each flow (hashed) and each INT metadata item, what was the previous sum. When a telemetry report arrives, it sums up incoming metadata values, compares them to the previous sums, and if the difference is greater than the threshold, the new sum is stored and the packet is marked to be sent to the INT monitor. Much like for the \emph{per-hop} algorithm, this is done through an unrolled loop. Finally, the moving average works similarly to the flow sum algorithm, however, it also applies a simple exponential moving average between the current and last received value. A configurable $\alpha$ parameter decides how much weight new values will have on the average, i.e., the smoothness of the function. Again, if the difference is greater than the threshold, the new value is stored, and the packet is marked. Finally, there is also a \emph{noop} algorithm, which always detects an event. 

If more complex event detection is required, it is possible to use \emph{complex\_detection} action for the algorithm table. Complex detection only takes a single parameter, which is a register table index. In this register, an expression according to the FastReact format is stored, and allows specifying more complex, stateful expressions, such as {\small \texttt{hop-latency > 10 and queue~buildup > 100}}. As such complex expression have significantly higher execution time due to complex logical operations, they will increase the per packet latency. 

We implemented our approach in P4\_16 language on the Netronome Agilio NFP-4000 platform~\cite{nfp4000}, which is a multi-threaded, multicore programmable network interface card. It distributes incoming packets among 48 packet processing cores, each running 8 threads. Memory is divided into a multi-tier structure, where tables and registers are stored in a \SI{2}{GiB} DRAM, utilizing two blocks of \SI{2}{MiB} SRAM as a cache. %The NFP4000 platform is programmed using firmware files, which is compiled from P4 sources files, through the Netronome Datapath Programming Tools. The compiled firmware file is then coupled with interface, register and table configuration files and installed on the network card. 
Our P4 implementation allows the SDN controller to dynamically reconfigure the INT behavior of each switch. This includes changing the switch role (sink, source or transit), and configuring which INT fields should by inserted by each switch, as well as changing the threshold and algorithm settings per flow. In our evaluation, we will only look at the \emph{fast\_detection}, and leave the \emph{complex\_detection} for future work.

\section{Evaluation and Results}

\begin{figure}
    \centering
    \includegraphics[width=\linewidth]{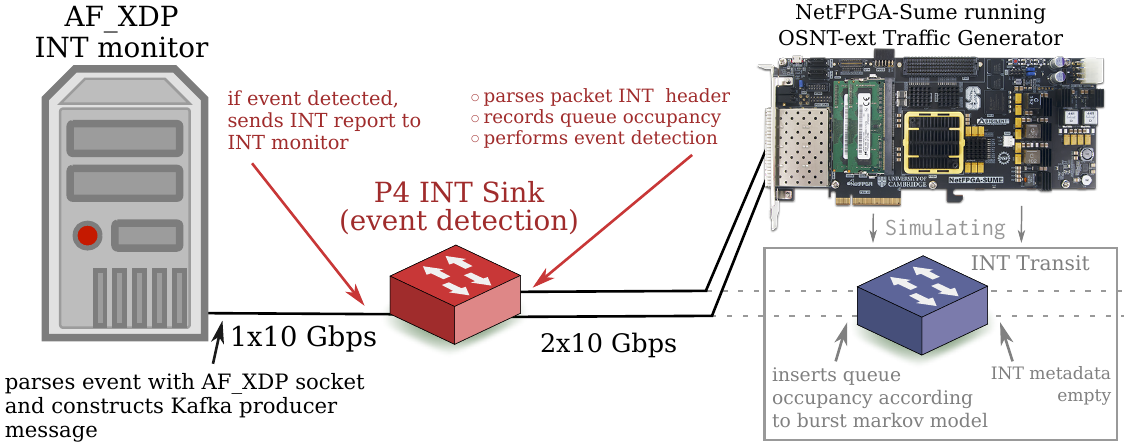}
    \caption{Testbed setup used for evaluation}
    \label{fig:testbed}
\end{figure}

%\todo[inline]{add a little bit research questions, like HULA paper.}
In this section we evaluate the  effectiveness of our approach by answering the following questions: 

\begin{enumerate}
    \item \textbf{INT monitor (Section \ref{sec:int-monitor}):} What is the performance capacity of the INT monitor, and how is this performance capacity affected by the telemetry item count and number of hops traversed by the INT report?
    \item \textbf{INT sink (Section \ref{sec:event}):} What is the potential processing capacity for the entire INT deployment, given different traffic patterns and event detection algorithms? %Also, how does this compare to using no event detection?
\end{enumerate}

Our testbed consists of an INT sink, an INT monitor and a traffic generator. 
The INT sink is connected to the traffic generator using 2x10 Gbps SFP+ connections, and is also connected using 1x10 Gbps SFP+ connector to the INT monitor. The INT sink is equipped with a Netronome Agilio 2x40G NFP-4000 SmartNIC, split into eight \SI{10}{Gbps} ports using a breakout module, running our P4 implementation for INT parsing and event detection. Here, INT reports incoming from the traffic generator (representing an INT-enabled network) are processed, and if an event is detected, an event report is sent to the INT monitor. 
The INT monitor is connected to the INT sink; processes incoming INT event reports, and parses them into a Kafka message. The constructed Kafka message is, however, not transmitted to a Kafka cluster, but is constructed and then dropped, since evaluating the Kafka message producer is out of scope of this paper. The event report parsing is done using an AF\_XDP~\cite{brouer18} socket. The machine is equipped with a 20 core Intel\textregistered~Xeon\textregistered~Silver 4114 CPU, running at \SI{2.20}{GHz}, and with \SI{32}{GB} RAM running Ubuntu~18.04~LTS. This end-host is also equipped with two 2x10G SFP+ Intel X520 network cards. 
Finally, the traffic generator is a NetFPGA-SUME\footnote{\url{https://github.com/NetFPGA/NetFPGA-SUME-public/wiki}} running the \emph{extmem}\footnote{\url{https://github.com/NetFPGA/OSNT-Public/wiki/OSNT-SUME-extmem-project}} variant of the Open Source Network Tester (OSNT)~\cite{antichi14}. This platform is equipped with four \SI{10}{Gbps} SFP+ interfaces, where OSNT-extmem can use two of them to generate synthetic traffic at \SI{20}{Gbps}. 
The entire testbed setup is shown in Figure~\ref{fig:testbed}. 

\subsection{INT Monitor}
\label{sec:int-monitor}
\begin{figure}
    \centering
    \includegraphics[width=\linewidth]{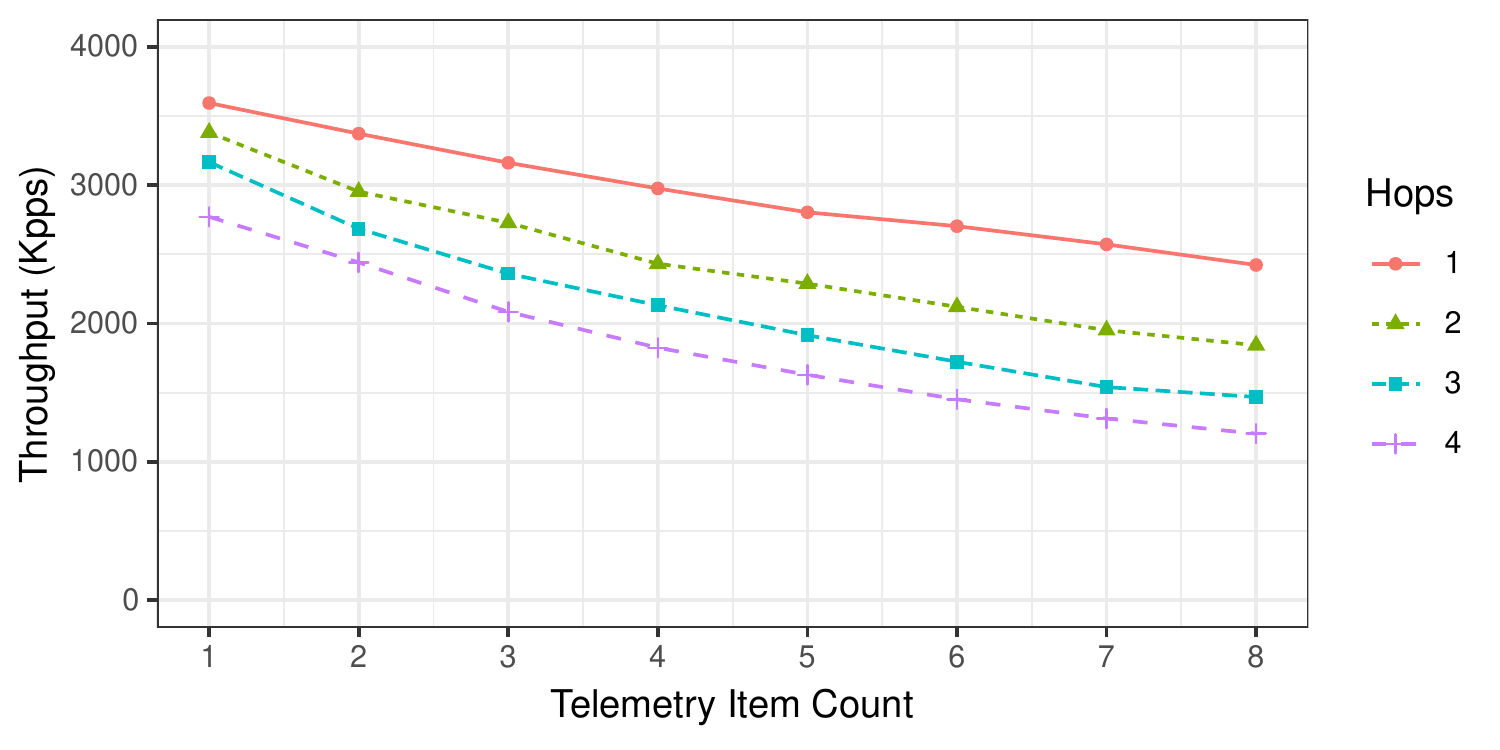}
    \caption{Event report processing capacity of the AF\_XDP INT monitor, per core/interface, varying header and hop count. }
    \label{fig:monitor}
\end{figure}

First, we evaluate the processing capacity of the INT monitor. %While our approach can reduce the INT traffic received by the INT monitor by several magnitudes, it is still interesting to see how our INT monitor implementation can cope with large networks, or networks which generate a large amount of INT events. To evaluate the processing capacity, 
We connected the INT monitor directly to the NetFPGA running the OSNT traffic generator, and generated \SI{10}{Gbps} of INT event report traffic for \SI{60}{\s}, while the monitor logged the number of events it was able to process per second. The result is shown in Figure~\ref{fig:monitor}, where the y-axis shows the number of packets processed per second and the x-axis shows the number of telemetry items included per hop in the event report. Different curves represent different number of hops in the topology, i.e., different number of switches from source to sink that appended telemetry items to the packet INT header. From the figure, we can see that the throughput depends on the size of the INT-event report: With small INT reports, containing only one telemetry item from a single switch, the INT monitor can process \SI{3593}{Kpps} of event reports. However, when the number of telemetry items increases to four, the processing capacity is reduced to \SI{2976}{Kpps} and the throughput to \SI{2771}{Kpps}. Finally, increasing the number of telemetry items to eight and the number of hops to four, which means each INT event report carries 32 values, reduces processing capacity to \SI{1204}{Kpps}. From these results, it is evident that the performance degradation due to increased number of hops is greater than that of due to increased number of telemetry items. This is because the INT parser needs to parse the entire INT instruction mask for each hop, regardless of how many entries are included. 

%1        1    1 3593.055   3.702660  0.4947888   0.9915789
%13       4    1 2975.637   1.234009  0.1579987   0.3160445
%4        1    4 2770.701 514.313727 66.3976166 132.8613240
%29       8    1 2422.644 273.375248 31.9961527  63.7831646
%32       8    4 1203.784 117.209418 12.6390166  25.1297510

\subsection{Event Detection}
\label{sec:event}

In this section, we evaluate our combined INT framework, including the event pre-filtering at the INT sink and the INT monitor. We vary the event detection algorithm and threshold setting. We connect the OSNT traffic generator to the INT sink, which in turn is connected to the INT monitor, as shown in Figure~\ref{fig:testbed}. In our evaluation, we look at the micro-burst event detection of a single INT-enabled switch. The traffic generated follows a burst Markov model~\cite{Zhang:2017:HMD:3131365.3131375}, which is based on micro-burst measurements taken from a Facebook data center, and provides the duration of and inter-burst time for each micro-burst. 
We emulate three different traffic patterns: \emph{web}, which is requests to and responses from web servers, \emph{cache}, which consists of traffic from in-memory caching servers used by the web servers and \emph{hadoop}, which is servers used for offline analysis and data mining. 

Three traces were generated, each consisting of $200,000$ packets, which emulate INT traffic by using the inter-burst times and durations provided in the model for one switch reporting its queue occupancy. Larger traces would have been preferable, however, the OSNT traffic generator, even with the \emph{extmem} extensions limits how many unique packets can be loaded into the NetFPGA memory. The traces are repeated for \SI{60}{\s}, continuously generating \SI{20}{Gbps} of traffic. From the recorded number of processed packets per second, and knowing the processing capacity of the INT monitor, we  calculate the potential INT report processing capacity per processing core of the monitor. For the threshold setting, we use \emph{queue occupancy} specified in microseconds. We have also included the threshold of \SI{0}{\us} in both experiments, which show the processing capacity if no event detection is done in the data plane. Here, the processing capacity is the same regardless of traffic model as all events are sent to the INT collector. 

\begin{figure}
    \centering
    \includegraphics[width=\linewidth]{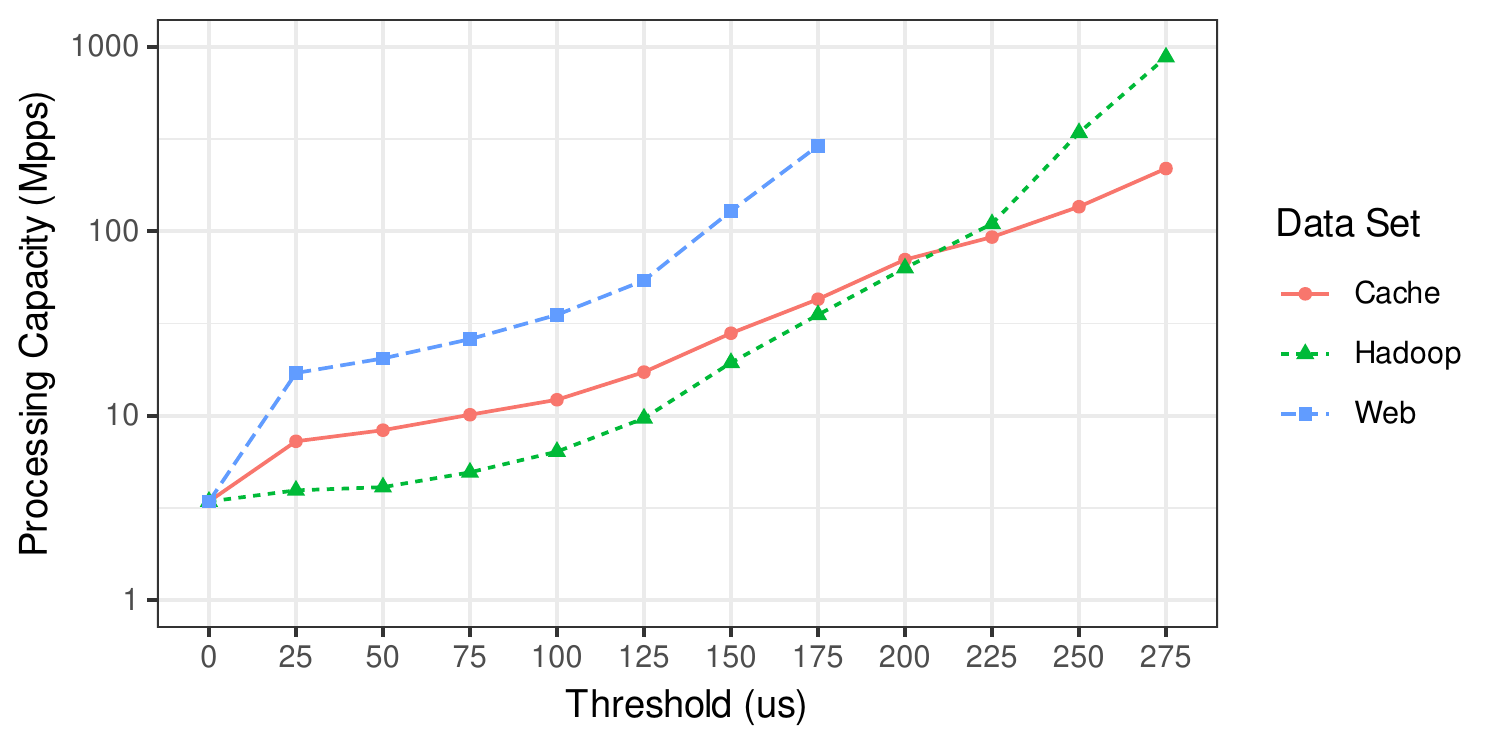}
    \caption{Processing capacity of the INT monitor using the \emph{per-hop/per-flow} threshold algorithm, with three different traffic types, varying the threshold setting. }
    \label{fig:thresh}
\end{figure}

%Jonathan: TODO: The numbers don't make 100% sense, the result form headers-1-hops-1 should be the same as markov-ref. Why is it not?
%double check experiment setup carefully, at least re-running headers experiment is fast. 
Results obtained by implementing the \emph{per-hop} or the \emph{per-flow} threshold algorithm are shown in Figure~\ref{fig:thresh}: The y-axis shows the potential processing capacity per processing core in the INT monitor in terms of packets per second (pps) including the pre-filtering in the INT sink, while the x-axis shows the configured threshold for the \emph{per-hop}/\emph{per-flow} algorithm. As we only generate one flow for these experiments, both algorithms have the same behavior. %Three different lines represent three different traffic patterns. 
We observe that performing no event filtering gives us a \SI{3.43}{Mpps} processing rate. With the \emph{web} traffic model, using a threshold of \SI{100}{\us}, the processing capacity reaches \SI{35.22}{Mpps}, and with a threshold of \SI{150}{\us} the processing capacity is \SI{128.59}{Mpps}. For the \emph{cache} traffic model, which has longer bursts and thus higher queue buildup, a threshold of \SI{100}{\us} results in \SI{12.21}{Mpps} processing capacity, while threshold of \SI{150}{\us} results in the perfornace of \SI{28.01}{Mpps}. 
As the \emph{web} traffic has lesser and shorter bursts of traffic, the number of events generated are fewer which gives a higher processing capacity, while more bursty traffic like \emph{cache} and \emph{hadoop} gain comparatively less from in-network event detection. 
The missing data points from the figure mean that no events were detected with that threshold. This is especially prevalent in the \emph{web} traffic model, as queue occupancy does not reach above \SI{175}{\us}. Given longer traces such events are, although rare, more likely to be present. It is important to note, however, that the selected settings are based on the Facebook traces and the NFP4000 card characteristics. Real networks with different queue sizes and link latencies, would require different thresold settings. 

\begin{figure}
    \centering
    \includegraphics[width=\linewidth]{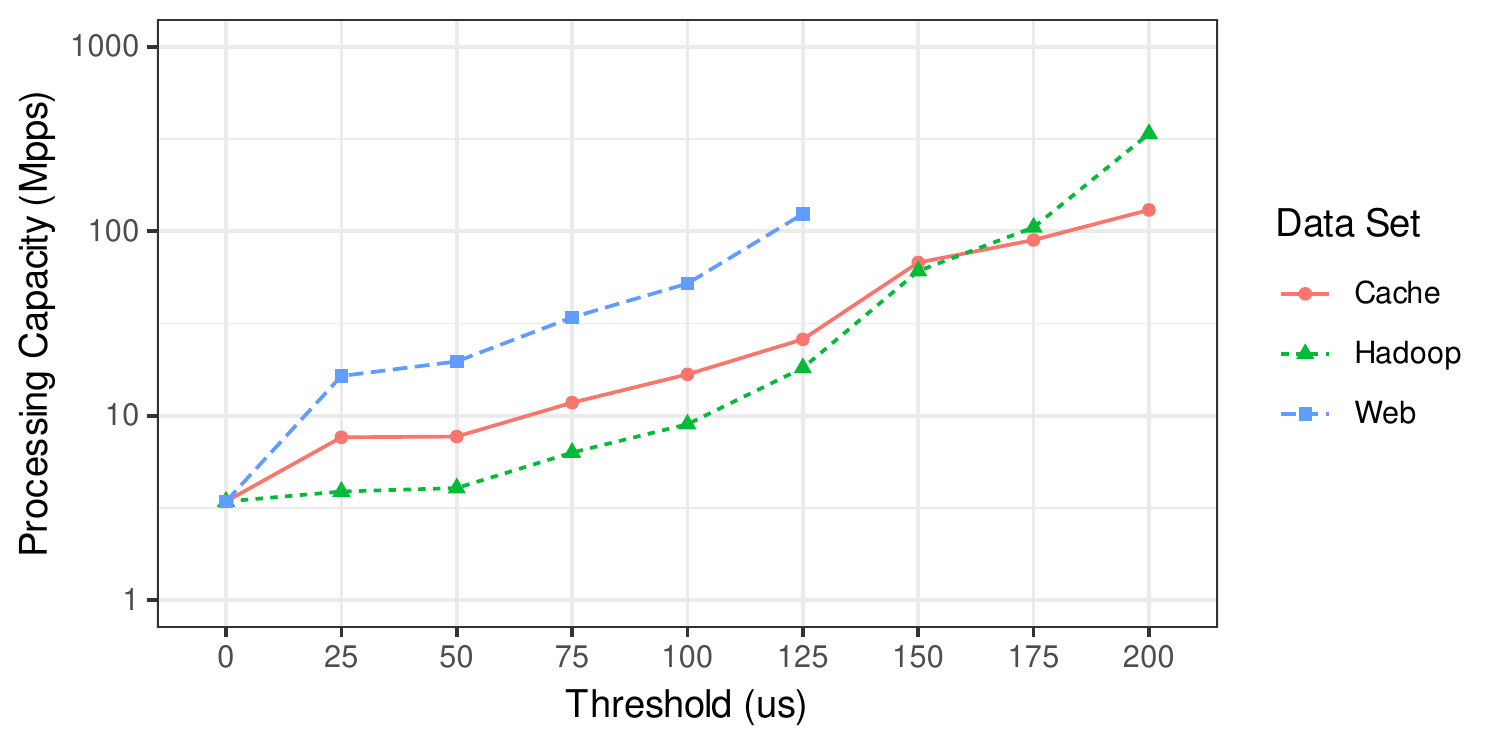}
    \caption{Processing capacity of the INT monitor using the \emph{moving average} threshold algorithm, with three different traffic types, varying the threshold setting. }
    \label{fig:avg}
\end{figure}

Finally, results from using the \emph{moving average} threshold algorithm can be seen in Figure~\ref{fig:avg}. As follows, with the \emph{web} traffic model and with a threshold of \SI{100}{\us}, the processing capacity is \SI{52.07}{Mpps}, and with a threshold of \SI{150}{\us} no events are detected. Using the \emph{cache} traffic model and a threshold of \SI{100}{\us}, the per-core processing capacity reaches \SI{16.76}{Mpps}, and with a threshold of \SI{150}{\us} it reaches \SI{67.64}{Mpps}. That is, we observe similar patterns as those observed for the \emph{per-hop}/\emph{per-flow} algorithm. However, moving average, in most cases, resulted in less data sent to the INT monitor, as shorter bursts are smoothed out, which allows increased total processing capacity of the INT-enabled network. The results demonstrate that our approach is highly scalable, and significantly reduces the network overhead and stream processor load by using effective event pre-filtering inside the switch in the data plane.

\section{Conclusions}
In this paper, we developed a programmable event detection mechanism for INT metadata. Our solution comprise two parts: First, P4-programmable switches use INT metadata to collect state information from the data plane. Using different filtering mechanisms, the amount of unimportant telemetry information that is sent to the stream processor is significantly reduced, something which reduces both the monitoring overhead and the load on the stream processor. The filtering can be programmed on a per-flow basis from the control plane. Second, our solution comprises a high-performance telemetry report parser in AF\_XDP, which streams remaining telemetry items to a distributed Kafka cluster and Elastic Search stack for further processing (e.g., analytics and machine learning) and visualization. In our evaluation, we use a testbed with P4-programmable network interface cards, and use several traces from real data-center workloads to evaluate the impact of different event detection algorithms and threshold configurations on the event detection ratio and load on the stream processor. The paper shows that our solution scales to process several hundred millions of telemetry headers per second. In our future work, we will extend the event detection algorithms, and evaluate the impact of the more complex event detection mechanisms on switch performance.

\section*{Acknowledgement}
Parts of this work have been funded by the Knowledge Foundation of Sweden (KKS) through the profile HITS.
\def\bibfont{\footnotesize}
\printbibliography

\end{document}